%Paper: hep-th/9410041
%From: CADONI@CAGLIARI.INFN.IT
%Date: Fri, 7 OCT 94 11:18 GMT

%%%%%%%%%%%%%%%%%%%%%%%%%%%%%%%%%%%%%%%%%%%%%%%%%%%%%%%%%%%%%%%%%%%%%%
%%                         Body of the paper                        %%
%%%%%%%%%%%%%%%%%%%%%%%%%%%%%%%%%%%%%%%%%%%%%%%%%%%%%%%%%%%%%%%%%%%%%%
%%			    Tex instructions			    %%
%%%%%%%%%%%%%%%%%%%%%%%%%%%%%%%%%%%%%%%%%%%%%%%%%%%%%%%%%%%%%%%%%%%%%%

\font\titolino=cmbx10
\font\tsnorm=cmr10
\font\tscors=cmti10

\font\tscorsp=cmti9
\magnification=1200
\def\ha{{1\over 2}}\def\efi{e^{-2\phi}}
\def\rrq{{r(r-\rq)}}
\def\mp{\left(1-{\rp\over r}\right)}\def\mq{\left(1-{\rq\over r}
\right)}
\def\e{{e}^}\def\ba{{a\Lambda}}\def\ab{{1\over a\Lambda}}
\def\iab{{\omega\over a\Lambda}}\def\iam{{\pi\omega\over 2a\Lambda}}
\def\r{r_*}\def\ub{\bar u}\def\vb{\bar v}
\def\nor{{1\over\sqrt{4\pi\omega}}}
\def\uno{^{(1)}}\def\due{^{(2)}}\def\norm{{1\over\sqrt{2\sinh\pia}}}
\def\und{^{(1)\dagger}}\def\dud{^{(2)\dagger}}
\def\pia{{2\pi\omega\over a\Lambda}}\def\uf{{\ ^F u}}
\def\up{{\ ^P u}}
\def\af{asymptotically flat }\def\ads{anti-de Sitter }
\def\bh{black hole }\def\st{spacetime }\def\el{extremal limit }
\def\ads{anti-de Sitter }\def\trans{transformation }
\def\bhs{black holes }
\def\ct{coordinate transformation }\def\arcth{{\rm arctanh}}

\def\sch{Schwarzschild }\def\coo{coordinates }\def\gs{ground state }
\def\Mi{Minkowski }\def\RN{Reissner-Nordstr\"om }
\def\br{back reaction }
\def\bc{boundary conditions }
\def\ds{ds^2=}\def\ef{e^{2(\phi-\phi_0)}}\def\F{F_\mn}

\def\rp{r_+}\def\rq{r_-}\def\qy{\left(1+{Q\over y}\right)}
\def\mn{{\mu\nu}}
\def\dd{\partial_+}\def\dm{\partial_-}
\def\sm{\sigma^-}\def\sp{\sigma^+}
\def\xm{x^-}\def\xn{x^+}\def\xp{x^+}\def\ma{\left(\xm - \xn\right)}
\def\af{asymptotically flat }\def\ads{anti-de Sitter }
\def\bh{black hole }\def\st{spacetime }\def\el{extremal limit }

\def\coo{coordinates }\def\gs{ground state }
\def\Mi{Minkowski }\def\RN{Reissner-Nordstr\"om }
\def\bc{boundary conditions }
\def\ds{ds^2=}\def\ef{e^{2(\phi-\phi_0)}}\def\F{F_\mn}

\def\ed{e^{-2\phi}}
\def\rp{r_+}\def\rq{r_-}\def\qy{\left(1+{Q\over y}\right)}
\def\ADS{$ADS^0$ }\def\PDS{$ADS^+$ }
\def\la{\Lambda}
\def \lq{\la^2}
\def\pho{e^{-2\phi_0}}

\def\2d{two-dimensional }
\def\3d{three-dimensional }
\def\4d{four-dimensional }

\def\PRD{{\tscors Phys. Rev. D }}
\def\PRL{{\tscors Phys. Rev. Lett. }}
\def\NPB{{\tscors Nucl. Phys. B }}

\def\PLB{{\tscors Phys. Lett. B }}

\def\MPLA{{\tscors Mod. Phys. Lett. A  }}
\def\CQG{{\tscors Class. Quantum Grav. }}

\def\note{\advance\notenumber by 1 \footnote{$^{\the\notenumber}$}}
\def\ref#1{\medskip\everypar={\hangindent 2\parindent}#1}
\def\beginref{\begingroup
\bigskip
\leftline{\titolino References.}
\nobreak\noindent}
\def\endref{\par\endgroup}
\def\beginsection #1. #2.
{\bigskip
\leftline{\titolino #1. #2.}
\nobreak\noindent}

\nopagenumbers
%%%%%%%%%%%%%%%%%%%%%%%%%%%%%%%%%%%%%%%%%%%%%%%%%%%%%%%%%%%%%%%%%%%%%%
%%			 End of instructions		            %%
%%%%%%%%%%%%%%%%%%%%%%%%%%%%%%%%%%%%%%%%%%%%%%%%%%%%%%%%%%%%%%%%%%%%%%
%%%%%%%%%%%%%%%%%%%%%%%%%%%%%%%%%%%%%%%%%%%%%%%%%%%%%%%%%%%%%%%%%%%%%%
%%			       Title	       		            %%
%%%%%%%%%%%%%%%%%%%%%%%%%%%%%%%%%%%%%%%%%%%%%%%%%%%%%%%%%%%%%%%%%%%%%%
\null
\vskip 5truemm
\rightline {INFNCA-TH-94-26}
\vskip 30truemm
\centerline{\titolino NON-SINGULAR FOUR-DIMENSIONAL BLACK HOLES AND}
\bigskip
\centerline{\titolino THE JACKIW-TEITELBOIM THEORY}
\vskip 15truemm
\centerline{\tsnorm Mariano Cadoni and Salvatore Mignemi}
\bigskip
\centerline{\tscorsp Dipartimento di Scienze Fisiche,}
\smallskip
\centerline{\tscorsp Universit\`a  di Cagliari, Italy}
\smallskip
\centerline {\tscorsp and}
\smallskip
\centerline{\tscorsp INFN, Sezione di Cagliari,}
\smallskip
\centerline{\tscorsp Via Ada Negri 18, I-09127 Cagliari, Italy.}

\vskip 40truemm
\centerline{\tsnorm ABSTRACT}
\smallskip
\begingroup\tsnorm\noindent
%\baselineskip=10truemm
%%%%%%%%%%%%%%%%%%%%%%%%%%%%%%%%%%%%%%%%%%%%%%%%%%%%%%%%%%%%%%%%%%%%%%
%%			      abstract		          	    %%
%%%%%%%%%%%%%%%%%%%%%%%%%%%%%%%%%%%%%%%%%%%%%%%%%%%%%%%%%%%%%%%%%%%%%%

A four-dimensional dilaton-gravity action whose spherical reduction
to two dimensions leads to  the Jackiw-Teitelboim theory is presented.
A nonsingular black hole solution of the theory is obtained and its
physical interpretation is discussed.
The classical and semiclassical properties of the solution and of its
\2d counterpart are analysed.
The \2d theory is also used to model the evaporation process of the
near-extremal \4d black hole.
We describe in detail the peculiarities of the \bh solutions,
in particular the purely topological nature of the Hawking radiation,
in the context of the Jackiw-Teitelboim theory.

%%%%%%%%%%%%%%%%%%%%%%%%%%%%%%%%%%%%%%%%%%%%%%%%%%%%%%%%%%%%%%%%%%%%%%
%%			   End of abstract			    %%
%%%%%%%%%%%%%%%%%%%%%%%%%%%%%%%%%%%%%%%%%%%%%%%%%%%%%%%%%%%%%%%%%%%%%%
%%%%%%%%%%%%%%%%%%%%%%%%%%%%%%%%%%%%%%%%%%%%%%%%%%%%%%%%%%%%%%%%%%%%%%
%%			       Address			            %%
%%%%%%%%%%%%%%%%%%%%%%%%%%%%%%%%%%%%%%%%%%%%%%%%%%%%%%%%%%%%%%%%%%%%%%
\vfill
\leftline{\tsnorm PACS: 97.60.Lf,04.60.+n,11.17.+y \hfill}
\smallskip
\hrule
\noindent
\leftline{E-Mail: CADONI@CA.INFN.IT\hfill}
\leftline{E-Mail: MIGNEMI@CA.INFN.IT\hfill}
%\bigskip
\endgroup
\vfill
\eject
\footline{\hfill\folio\hfill}
\pageno=1
\baselineskip=18pt
%%%%%%%%%%%%%%%%%%%%%%%%%%%%%%%%%%%%%%%%%%%%%%%%%%%%%%%%%%%%%%%%%%%%%%
%%				text                      	    %%
%%%%%%%%%%%%%%%%%%%%%%%%%%%%%%%%%%%%%%%%%%%%%%%%%%%%%%%%%%%%%%%%%%%%%%
\beginsection 1. Introduction.
In the last years dilaton-gravity theories have been widely
investigated both in two and four spacetime dimensions [1-6].
In particular the model proposed by Callan, Giddings, Harvey and
Strominger (CGHS) [1] has generated new interest and activity on
black hole physics.
It was recognised that \2d dilaton-gravity models represent a
theoretical
laboratory for investigating the fundamental issue of information loss
in the \bh evaporation process.
Moreover the CGHS model can be viewed as the low-energy effective
theory that describes the S-wave sector in the evaporation process of
the near-extreme magnetically charged dilaton \bh in four dimensions.
The \4d \bh is also of interest because it is a classical solution
of low-energy effective string field theory [3,4].

In a previous paper [5] we found \4d dilaton \bh solutions
which can be considered as a generalization of the
Garfinkle-Horowitz-Strominger (GHS) solutions [3,4].
They are also classical solutions of a
field theory that arises as a low-energy approximation to string
theory.
The corresponding \2d effective theory  that describes  a magnetically
charged \4d \bh near its extremality represent a generalization
of the CGHS model [6].
In this paper we will study in detail another special case of these
models,
namely the one whose dimensional reduction to two dimensions leads to
the Jackiw-Teitelboim (JT) theory [7].
The relevance of this model is twofold. From a purely \4d point of
view the corresponding magnetically charged \bh solutions evidenciate
properties which are in some sense intermediate between the \RN (RN)
and the GHS black holes and in addition enjoy
the peculiar property of being free from curvature singularities.
On the other hand the spherical reduction to two dimensions leads
to the JT theory. The latter can  therefore be used to model the
S-wave scattering
of the 4d \bh near extremality.
In this context the dynamics of the JT theory is very peculiar.
In fact the \bh solutions of the JT theory present features which have
no correspondence in other theories.  Their existence in the
context of the theory is related to the choice of the boundary
conditions
for the spacetime, they have in some sense a purely topological
origin.
As a consequence the  Hawking evaporation process is a purely
topological
effect. Our results are also interesting with respect to the
\bh solutions in three dimensions. In fact has been demonstrated [8]
that the JT theory arises from dimensional reduction of
the  \bh solutions of Ba\~nados, Teitelboim and Zanelli (BTZ) [9].
Thus our results about the black hole physics in the JT theory
can be as well used to model the evaporation process of a
\3d \bh.

The structure of the paper is as follows. In sect. 2 we study the \4d
model. In particular we derive the magnetically charged \bh solutions,
we study their local and global properties and their behaviour
in the extremal limit. In sect. 3 we consider the \2d theory obtained
by spherical reduction of the \4d one. The classical properties of the
corresponding \bh solutions are analysed at length in particular in
connection with  the \4d ones. In sect. 4 we study the semiclassical
behaviour of the \2d theory, in particular the evaporation process of
the \bh including the back reaction.
The main results of our investigation
are summarized in sect 5.
\beginsection 2. The \4d  model.
Let us consider the 4d action:
$$S=\int\sqrt g\ d^4x\ \efi(R-F^2),\eqno(2.1)$$
where $\F$ describes a Maxwell field and $\phi$ a scalar (dilaton).

The relevance of this action resides in the fact that its
dimensional reduction to two dimensions leads, as we shall see
in the following,
to the Jackiw-Teitelboim theory. Moreover, it is a special case
($k=0$)
of the low energy effective string actions discussed in [5,6].
A conformal transformation of the metric field
$\hat g_\mn=e^{-2\phi}g_\mn$
leads to the minimally coupled action:
$$S=\int\sqrt{\hat g}\ d^4x\ [\hat R-6(\hat\nabla\phi)^2-\efi F^2].
\eqno(2.2)$$
However, we prefer to discuss the action in the form (2.1).
One of the reasons
is that the solutions of (2.1) describe non-singular black holes
similar to those discussed  in [8,10] for the JT theory.

The field equations stemming from (2.1) are:
$$R_\mn-\ha g_\mn R=2F_{\mu \rho}F^{\rho}_{\nu}-\ha g_\mn F^2+
e^{2\phi}\bigl[(\nabla_\mu\nabla_\nu-
g_\mn\nabla^2)\efi\bigr],$$
$$R=F^2,\eqno(2.3)$$
$$\nabla_\mu(\efi F^\mn)=0.$$
A spherically symmetric \bh solution of the field equation is
given [6] by a magnetic monopole:
$$F_{ij}={Q_M\over r^2}\epsilon_{ij},$$
with metric
$$\ds -\mp dt^2+\mp^{-1}\mq^{-1} dr^2+r^2 d\Omega^2\eqno(2.4)$$
and dilaton field
$$\ef=\mq^{-1/2}.\eqno(2.5)$$
The two parameters $r_+$ and $r_-$ ($ r_+\ge r_-$) are related to the
charge $Q_M$ and
to the mass $M$ of the  \bh by the relations:
$$2M=r_+,\qquad\qquad Q_M^2={3\over 4}r_+r_-.\eqno(2.6)$$

The temperature and entropy of the solution are given respectively
by [6]:
$$T={1\over 4\pi\rp}\left(1-{\rq\over\rp}\right)^{1/2}=
{(M^2-{Q_M^2/3})^{1/2}\over 8\pi M^2},\qquad\qquad
S=\pi r_+^2=4\pi M^2.\eqno(2.7)$$
The temperature vanishes in the extremal limit $\rp=\rq$, i.e.
$M^2={1\over3}Q_M^2$, which
should therefore be considered as the ground state for the Hawking
evaporation process of
a \bh of given charge.

The spatial sections of the metric coincide with those of the \RN
solution of general relativity and those of the
Garfinkle-Horowitz-Strominger
solution of the low-energy effective action of string theory [3,4].
Owing to the difference in the $g_{00}$ component, however,
the three metrics possess quite different physical properties.

The metric (2.4) is \af and the curvature is regular everywhere
except at $r=0$.
This point, however, is not part of the manifold, since the range
of the radial coordinate is given by $r\ge\rq$. In fact, even if
the manifold is
regular at $r=\rq$ (a coordinate singularity is placed at this point,
but the curvature tensor is regular), for $r<\rq$
the metric becomes euclidean and the dilaton imaginary, so that one is
forced to cut the manifold at $r=\rq$.
Furthermore, a horizon is present at $r=\rp$. Hence,
the solution (2.4) describes a
non-singular four-dimensional \bh. This is reminiscent of the regular
two-dimensional \bh discussed in [8,10].
For negative values of $\rp$ and $\rq$,
instead, one has negative mass and a naked singularity.

A better understanding of the causal structure of the spacetime
can be obtained by discussing
its  maximal extension and the Penrose diagram.
Let us thus consider only the \2d
$r-t$ sections of the metric and introduce the
"Regge-Wheeler tortoise" coordinate, defined by:
$$d\r=\mp^{-1}\mq^{-\ha}dr.\eqno(2.8)$$
The change of variables $u=t-\r$, $v=t+\r$ takes  the metric in
the form
$$\ds -\mp dudv,\eqno(2.9)$$
with $r$ defined implicitly in terms of $u$ and $v$ by (2.8).
In these \coo the metric is clearly regular at $r=\rq$.
One can now perform another change of coordinates which eliminates
the singularity at $r=\rp$:
$$U=-\exp (-\beta u),\qquad\qquad
V=\exp (\beta v),$$
where $\beta= \ha\sqrt{\rp-\rq\over\rp^3}$.
The final result is the metric expressed in the Kruskal form:
$$\eqalign{\ds- {(\rp)^{-(1+\gamma/2)}\over\beta^2 r}
e^{-2\beta\sqrt{r(r-\rq)}}
&\left(\sqrt{r-\rq}+\sqrt{r}\right)^\gamma\cr
\times&\left(\sqrt{\rp(r-\rq)}+\sqrt{r(\rp-\rq)}\right)^2dUdV\cr}
\eqno(2.10)$$
with
$$\gamma=-2\beta (2\rp+\rq),\qquad\qquad
UV=-\exp(2\beta \r),\qquad\qquad
{V\over U}=-\exp(2\beta t).$$

This form of the metric is regular both at $\rp$ and $\rq$.
In a standard way
one can deduce from it the form of the Penrose diagram (fig.1),
which results analogous to that of the \sch solution. The only
difference is that the singularity is now replaced by
a inner horizon, which represents the boundary of the manifold.

In the extremal case one can proceed in a similar way. However,
we have not been able to find an explicit form for the Kruskal
metric. The Penrose diagram is given in fig. 2 and is identical
to that of the GHS solution.

In the following, we shall be especially interested in the
properties of the
extremal limit $\rp=\rq$. For this purpose is useful to define a
new coordinate $\eta$, such
that $\eta={\rm arcsh}\sqrt{r-\rp\over\rp-\rq}$.
In terms of $\eta$, the metric and the dilaton field (2.4),
(2.5) take the form
$$\eqalign{\ds -4Q^2&{\Delta\sinh^2\eta\over\rp+\Delta\sinh^2
\eta}dt^2+(\rp+\Delta\sinh^2\eta)^2(4d\eta^2+d\Omega^2),\cr
&\ef=\left[{\rp+\Delta\sinh^2\eta\over\Delta\cosh^2\eta}\right]^\ha,
\cr}
\eqno(2.11)$$
where $\Delta=\rp-\rq$ and $Q=(2/\sqrt3) Q_M$.

In the limit $\Delta\to 0$, the spatial sections of our
solution can be described as an \af
region attached to an infinitely long tube (the "throat"),
in analogy with
the RN and GHS case. It is not possible, however, to describe the
\el by a unique metric: rather, there are several regimes under
which the limit can be approached, which
correspond to different solutions of the action (2.1).

The region of the throat near the horizon is described, for
positive $\rp$ and $\rq$, by the solution:
$$\eqalign{\ds-4Q^2&\sinh^2\eta\ dt^2+Q^2(4d\eta^2+d\Omega^2),\cr
&\ef={Q\over\cosh\eta}.\cr}\eqno(2.12)$$
In this limit the metric is the direct product of a 2d \st of
constant negative curvature and a two-sphere of radius $Q$.

The extremal limit of the negative mass solution,
with negative $\rp$ and $\rq$ and naked singularity is given
instead by:
$$\eqalign{\ds-4Q^2&\cosh^2\eta\ dt^2+Q^2(4d\eta^2+d\Omega^2),\cr
&\ef={Q\over\sinh\eta},\cr}\eqno(2.13)$$
It appears that in the extremal limit also the negative mass solution
becomes regular:
the metric (2.13), in fact, is the direct product of a 2d \st of
constant negative curvature and a two-sphere. We shall discuss in
more detail
the properties of this solution in the following.

Both solutions (2.12) and (2.13) tend, for $\eta\to\infty$ to
the metric
$$\ds-4Q^2e^{2\eta}dt^2+Q^2(4d\eta^2+d\Omega^2),\eqno(2.14)$$
with linear dilaton:
$$\phi=-\ha\eta,$$
which is a direct product of 2d \ads spacetime and a two-sphere and
corresponds to the throat region.

Finally, the \af region and the throat can be described by the
solution:
$$\eqalign{\ds-4Q^2&\qy^{-1}dt^2+\qy^2(dy^2+y^2d\Omega^2),\cr
&\ef=\qy^\ha\cr}\eqno(2.15)$$
and $y=r+Q\ge 0$. This metric is everywhere regular and describes the
transition
between an \af \st for $y\to\infty$ and one with topology
$H^2\times S^2$ for
$y\to 0$, $H^2$ being 2d \ads \st.

An unpleasant property of the solution (2.4) is that, contrary for
example
to the GHS solution,
it is not geodesically complete even in the extremal limit.
This is easily
seen by considering the geodesic equation for the radial motion:
$$\left({dr\over d\lambda}\right)^2=\mq\left[E^2-\mp\epsilon\right],
\eqno(2.16)$$
where $\epsilon =0,1,-1$ for light-like, time-like,
space-like geodesics
respectively, $E$ is the energy of the orbit and $\lambda$ is the
affine
parameter. The boundary at $r=\rq$ is in general at
finite  distance, but becomes in the extremal case infinitely far
away along
time-like and space-like geodesics.
In spite of this, light-like geodesics have finite length also in
the extremal
limit. In fact, for $\epsilon=0$, the geodesic length is given by
$$\lambda={1\over E}\left[\sqrt\rrq+\rq\ln\left(\sqrt{r-\rq}
+\sqrt r\right)\right],\eqno(2.17)$$
which is always finite for $r\ge\rq$.
The length of time-like geodesics, instead, is finite for $\rp>\rq$,
but tends to infinity in the extremal limit.

This should be compared with the RN and GHS solutions, which, as
already
remarked, have the same spatial sections as our solution.
In the extremal
RN metric, the horizon is at finite distance both along time-like and
light-like paths, while in the extremal GHS metric the distance is
infinite in
both cases. Space-like geodesic have instead infinite length for
RN and GHS extremal black holes. Our solution is therefore in some
sense intermediate between the two.
The lack of geodesically completeness of our solution can be however
remedied if one glues to the boundary of the spacetime at $r=\rq$
another copy of the same manifold. This is
possible because the extrinsic curvature vanishes there.

\smallskip
\beginsection 3. The  \2d model.
\smallskip
The two-dimensional action of the JT theory can be obtained by
spherical reduction of the \4d action (2.1) for a near-extreme
magnetically charged
\bh solution. This dimensional reduction has been described in [6]
in the context of a general \4d dilaton gravity theory of which the
action (2.1) represent just a special case.
{}From (2.1), taking  the angular coordinates to span a  2-sphere of
constant radius $Q$, we get the dimensionally reduced action:
$$S={1\over2\pi}\int\sqrt g\ d^2x\ e^{-2\phi}\left[R+2\lq\right].
\eqno(3.1)$$
where  $\la$ is related to the \4d magnetic charge by $\la=(2Q)^{-1}$.
The action (3.1) describes the JT theory [7].

The general time-independent solution of the corresponding field
equations is  by now
well-known and has been discussed at length in the
literature [11,6,8,10].
In \sch coordinates it has the form:
 $$\ds-(\la^2r^2-a^2)dt^2+(\la^2r^2-a^2)^{-1}dr^2,\qquad\qquad
\ef=(\la r)^{-1}.\eqno(3.2)$$
where $a^2$ is an integration constant which can assume both
positive and
negative values and is related to the Arnowitt-Deser-Misner (ADM)
mass of the solution by
$$M={1\over 2} \pho a^2\la. \eqno (3.3)$$
The ADM mass is calculated with respect to the asymptotic solution
with $a=0$.
Independently of the value of the parameter $a^2$
the solution (3.2) describes a spacetime of constant negative
curvature
$R=-2\lq$, i.e \2d anti-de Sitter space. The solutions (3.2)
with $a^2$
positive, negative or zero describe in \sch \coo the \2d section
of the extremal \4d solutions (2.12), (2.13), (2.14) respectively [6].
For $a^2>0$ the spacetime
(3.2) has an horizon at $r=a/\la$ indicating that it can be
interpreted
as a \2d \bh, whereas for $a^2<0$ the solution has negative ADM
mass\footnote{$^1$}{The negative mass is related to the choice of
the ground state which we have identified with the solution
with $a=0$. The reason for this choice will be explained later
on this section.}.
However the metrics with different values of the parameter $a^2$
represent different parametrizations of the same manifold, namely
anti-de Sitter spacetime, with coordinate patches covering
different regions of the space. This fact was first demonstrated,
for the metrics with $a^2>0$, in [8] (see also [6]).
Indeed one can easily find the coordinate transformations relating
the different metrics. The transformation
$$\eqalign{r &\to a\la t r,\cr e^{2a\la t}&\to \lq t^2 -
\la^{-2}r^{-2},\cr}\eqno(3.4)$$
brings the metric with $a^2>0$ into the metric with $a=0$,
whereas
$$\eqalign{r&\to {a\over2}\left(\la^{-2}r^{-1}-\lq r t^2- r\right),
\cr
\tan(a\la t)&\to {1\over 2}\left(\la t - \la^{-3}r^{-2}t^{-1}
-\la^{-1}t^{-1}\right),\cr}\eqno(3.5)$$
relates the metric with $a^2<0$ to the metric with $a=0$.
Notice that even though the solutions (3.2) locally describe the same
spacetime independently of the value of $a^2$, the expression for the
dilaton becomes  different for the three
cases after using (3.4) and (3.5).
To distinguish  different solutions in  (3.2) we will denote
with $ADS^+$, $ADS^0$ and $ADS^-$ the spacetimes corresponding to
positive, zero and negative $a^2$ respectively.

The equivalence of the metric part of the solutions (3.2) with
different values of $a^2$ up to
a coordinate transformation makes it difficult to interpret $ADS^+$
as a black hole. In fact the spacetime can be extended beyond $r=0$,
and the maximally extended spacetime is the whole of
\2d  anti-de Sitter spacetime which of course has no horizons
and is geodesically complete [8].
However $ADS^+$ can represent a \bh if we cut the spacetime off
at $r=0$.
The reason why one has to cut off the spacetime  at this point is
clear
if one takes into account the expression (3.2) for the dilaton.
By analytically continuing the spacetime beyond $r=0$ one would enter
in a region where $\exp(-2\phi)$ becomes negative. The \2d action
(3.1)
has been obtained by spherical reduction of the \4d action (2.1).
When this reduction is carried out, the area of the transverse sphere
of constant radius in four dimensions
becomes the factor $\exp(-2\phi)$ multiplying
the Ricci scalar in (3.1) and has therefore to be positive.
Thus if one wants to model a \4d near-extremal magnetic \bh
by means of a \2d solution of the action (3.1) one has to cut the
spacetime
off at $r=0$. The Penrose diagram for $ADS^+$ and $ADS^0$ are shown
in fig. 3 and 4.
Notice the very peculiar role played by the dilaton in the context of
the JT theory: it sets the boundary conditions on the spacetime,
making  solutions which have the same local properties topologically
not equivalent.

Once one has recognised $ADS^+$ as a \bh one can use it to model the
S-wave sector of the evaporation process of a \4d \bh
near its extremality. In particular one can associate to it
thermodynamical
parameters. Using standard formulae we have for the temperature
and entropy
of the hole:
$$\eqalign{T&={a\la\over 2\pi}={1\over2\pi}e^{\phi_0}
(2M\la)^{1/2},\cr
S&=4\pi e^{-\phi_0}(M/2\la)^{1/2}=2\pi e^{-2\phi_h},\cr}
\eqno(3.6)$$
where $\phi_h$ is the value of the dilaton at the horizon.
The specific heat of the hole is positive indicating that loosing
mass through the Hawking radiation the hole will set down to its
ground state which we have identified with $ADS^0$.

At this stage a careful analysis of the mass spectrum of the
solutions (3.2)
is necessary particularly in view of the discussion of the Hawking
evaporation
process which will be the subject of the next section.
The mass spectrum is labelled by a continuous parameter $M$,
which in principle
can be an arbitrary real number. From a \2d point of view there is
no reason to exclude the states with $M<0$. In fact, differently
from e.g.
\sch \bh in 4d, the states with $M<0$ do not correspond to naked
singularities of the spacetime. The metrics (3.2) describe
spacetime with
constant curvature for every value of $a^2$. The separation of
the spectrum
in states with positive and negative mass is a consequence of the
choice
of $ADS^0$ as the ground state. This choice is from a \2d
point of view again arbitrary, since the mass spectrum is in
principle unbounded and the system has no ground state.

It is interesting to compare this situation with that of the \3d BTZ
black holes [8,9].
The \3d \bhs (with angular momentum $J=0$) have a continuous mass
spectrum for $M>0$ and the vacuum
is regarded as the
empty space obtained by letting the horizon size go to zero:
$$ds^2_{vac} =- (\la r)^2 dt^2+(\la r)^{-2}dr^2 +r^2 d\theta^2.
\eqno(3.7)$$
For $M<0$ one has solutions which describe naked singularities, unless
$M=-1$. In this case the metric describes the \3d \ads space :
$$ds^2 =-[ (\la r)^2+1] dt^2+[(\la r)^2+1]^{-1}dr^2 +r^2 d\theta^2,
 \eqno(3.8)$$
One sees that the \3d \ads space emerges as bound state, separated
from the continuous
\bh spectrum by a mass gap of one unit, i.e. by a sequence of naked
singularities
which of course cannot be included in the  configuration space.
The dimensional reduction from three to two dimensions not only
makes all spacetimes of constant negative curvature locally
equivalent up
to a coordinate transformation, but also eliminates the mass gap
in the
spectrum making it continuous and unbounded.

Let us now explain how one can single out $ADS^0$ as the physical
ground state for the \2d theory. The point is again the
relationship between
the \2d theory and the \4d one described in the previous section.
In fact  the \2d \bh mass $ M$ is related to the parameter $\Delta$
in eq. (2.11) which measures the deviation from extremality of the
\4d \bh.
For $M<0$ we have $\Delta <0$ with negative $r_-$ and $r_+$, the
singularity
at $r=0$
becomes visible and the solutions (2.4) describe naked singularities.
Therefore if we want our two-dimensional theory to describe
nearly extreme \4d black holes  we have to consider only
the $M\ge 0$ range as the physical mass spectrum for our \2d \bh.

Let us now study the response of our \2d dilaton-gravity system
to the introduction
of matter. Consider $N$ massless fields $f_i$ conformally coupled
to the
2d-gravity model defined  by the action (3.1). The classical action
is:
$$S={1\over2\pi}\int d^2x\sqrt g\left[e^{-2\phi}\bigl(R+2\la\bigr)
-{1\over2}\sum_{i=1}^N(\nabla f_i)^2\right].\eqno(3.9)$$
In the conformal gauge
$$ds^2=-e^{2\rho}d\xn d\xm,\qquad x^\pm=x^0\pm x^1,  $$
the equation of motion and the constraints are
$$\eqalign{\dd\dm f_i&=0,\cr \dd\dm\rho&=-{\la^2\over4}e^{2\rho},\cr
\dd\dm\ed&= -{\la^2\over2}e^{2(\rho-\phi)},\cr
 \dd^2\ed &-2\dd\rho\dd\ed=-{1\over 2}\sum_{i=1}^N\dd f_i\dd f_i,\cr
 \dm^2\ed &-2\dm\rho\dm\ed=-{1\over 2}\sum_{i=1}^N\dd f_i\dd f_i.
\cr}\eqno(3.10)$$

The peculiarity of this system of differential equations is that
the equation which determines the metric (the conformal factor $\rho$)
is independent of both the dilaton and the matter fields.
As a consequence one can solve independently the equation for $\rho$,
the dilaton being determined afterwards by the expression for the
metric
and, through the constraints,  by the stress energy tensor of the
matter. This analysis has been carried out in [6], where it was
demonstrated that the vacuum solutions ($f_i=0$) of (3.10) can be
written,
using the residual coordinate invariance within the conformal
gauge, as
$$e^{2\rho}={4\over\la^2}\ma^{-2},\eqno(3.11)$$
$$\ef={\la\over 2}\ma,\eqno(3.12)$$
which represent $ADS^0$ in the conformal gauge. Here and in the
following we will consider only the region $\xm\ge \xp$ of the
spacetime which corresponds to the $r\ge 0$ region in \sch
coordinates.
The solutions generated  by a $f$-shock wave with stress tensor
$$T_{++}=2\alpha^2\la^{-1}e^{-2\phi_0}\delta(\xn-\xn_0),\eqno
(3.13)$$
are instead
$$e^{2\rho}={4\over\la^2} \ma^{-2},\eqno(3.14)$$
$$\ef={\la\over 2}{\ma\over 1-\alpha^2\xn\xm}.\eqno(3.15)$$
The constant $\alpha$ is related to the ADM mass of the solution by
$M=e^{-2\phi_0} \la^{-1}\alpha^2/4$.

The effect of the shock wave on the vacuum solution (3.11),(3.12) is
therefore encoded in the modification (3.15) of the dilaton,
the metric part (3.14)
of the solution being insensitive to the presence of matter.
However this is true only if one chooses to take a maximal extension
of the spacetime (3.14), i.e. if one continues analytically the
solution beyond the line $1-\alpha^2x^{+}x^{-}=0$ (the point $r=0$
in \sch coordinates), where $\exp(-2\phi)$ becomes negative.
As we have seen previously the interpretation of $ADS^+$ as a \bh
can be established only if one cuts off the spacetime at this line.
Therefore even though the local properties of the metric
 (3.14) are insensitive to the presence of matter, the global ones
(the topology) are not; the effect of the shock wave on the
vacuum (3.11),(3.12)
is to create a boundary of the spacetime along the line where
$\exp(-2\phi)$ becomes negative.

The analysis of the dynamical evolution of our \2d system in this
general setting could be implemented by imposing some appropriate
boundary conditions for the fields and then
by considering the dynamics of the boundary.
In this paper we are mainly interested in the semiclassical properties
of the \bh solution  such as the Hawking evaporation process:
it turns out that for the study of such process an exact knowledge
of the dynamics of the boundary is quite unnecessary.
In fact, since the Hawking evaporation is a process which takes place
outside the event horizon, to study it is enough to consider a
coordinate system in which the boundary at $r=0$ is not visible.
This can be done in a standard way starting from eq (3.2), defining
the "Regge-Wheeler tortoise" coordinate $r_*$ and light-cone
coordinates $\sigma^+,\sigma^-$ as follows:
$$r_*=-{1\over a\la}\arcth\left({\la r\over a}\right),\qquad
\sigma^+=t+r_*,\qquad\sigma^-=t-r_*.\eqno(3.16)$$
In these \coo, the solution (3.2) with $a^2>0$ becomes
$$\eqalign{ds^2&=-a^2 \sinh^{-2}\left [\left(\sigma^--\sigma^+\
\right){a\la
\over 2}\right],\cr
\ef&={1\over a}\tanh\left[\left(\sigma^--\sigma^+\right)
{a\la\over 2}\right].\cr}\eqno(3.17)$$
which represents $ADS^+$ in the conformal gauge.
The new coordinate system covers only the region $r>r_h$ of the
spacetime
defined by (3.2). The metric part of the solution (3.17) can be
brought into the form (3.14) by the transformation
$$\eqalign {x^+&={2\over a\la}\tanh\left(a\la\sigma^+/2\right)\cr
x^-&={2\over a\la}\tanh\left(a\la\sigma^-/2\right).\cr}\eqno(3.18)$$
{}From the previous equations one easily sees that the coordinates
$\sigma^+,\sigma^-$ cover only the region $\{-2/a\la<x^+<2/a\la,
-2/a\la<x^-<2/a\la\}$ of the spacetime defined by (3.11) which
represent $ADS^0$ in the conformal gauge.

The coordinate transformation  (3.18) gives the relationship between
the coordinate system
"free-falling"  on the horizon of the \bh
and the  "anti-de Sitter" asymptotic
coordinate system. This relationship is similar
to that between a Rindler and a Minkowski coordinate system
in \2d flat spacetime even though in our case the coordinate
transformation
does not correspond to the motion of any physical observer.
Moreover  in our case   the presence of the boundary at $r=0$,
which makes
$ADS^+$ and $ADS^0$ topologically not equivalent, has a dynamical
interpretation
in terms of the matter distribution in the spacetime.
In this sense the JT theory  has  a very interesting and particular
status in the context of the gravity theories in arbitrary
spacetime dimensions. In fact the distribution of matter does not
determine the local geometry of the space, as usually happens for
gravity theories,  but its global properties, i.e. its topology.
At first sight this conclusion seems to hold only for the
conformal  matter-gravity coupling defined by the action (3.9).
A more general coupling, for example a dilaton-dependent coupling
of the matter, could spoil this property. In fact as a result of such
a coupling  a direct relationship between the local geometry and the
distribution of matter would be at hand.
However such a coupling cannot have any impact on the feature
which seems
to be responsible for the particular  behavior of the theory,
namely the fact
that the solutions of the field equation in absence of matter describe
different parametrizations of the same space and can be therefore
physically distinguished only through boundary conditions.

The features of our model are in some sense shared by all the
\2d dilaton
gravity models. In fact it has been demonstrated that all the models
of this type are physically equivalent to a model of free matter
fields
reflecting off a dynamical moving mirror [12].
In this context the JT theory represents the extreme case: the
formulation of the model in terms of a dynamical boundary
is the only possible because the dynamics in terms of the
gravitational
and dilaton field is in some sense trivial.

\beginsection 4. Semiclassical properties.
As discussed before, the difference between the ground state and the
\bh \st is essentially of topological nature. As is well known,
different topologies of spacetimes with the same geometrical
background
give rise to different vacuum states for quantum fields and by
consequence the vacuum in one \st will be perceived as a thermal bath
of radiation by an observer in the other system.

A typical example of this is the quantization of scalar fields in a
Rindler spacetime. This \st describes flat space as seen by a
uniformly
accelerated observer. A transformation of \coo exists which puts the
Rindler line element into the \Mi form: however the \coo so defined
do not cover the whole \Mi\st (physically, an accelerated observer
cannot see the whole \st, but a horizon hides part of it to
his view).
It follows that two different Hilbert spaces are necessary for the
quantization of fields in Rindler and Minkowski spaces: the vacuum
state in one \st will be perceived as a thermal state in the other.

Our case is analogous: the \PDS \bh is related by a \ct to the \ADS
ground state, but its image does not cover the whole \ADS: for this
reason it is not possible to define the same vacuum state for the two
spacetimes and a thermal radiation will be observed.

Two main differences are however present in our case: first of all,
the \ct between \PDS and \ADS does not correspond to the motion of
any physical
observer, so that the two \st should be considered as physically
distinct,
and not as the same \st  seen from different observers. Second, the
\ads \st has a large group of symmetries, so that the \ct (3.18)
between (3.11) and (3.17) is not uniquely defined, as we shall see
in the
following. Of course, this fact has no consequence on the physical
results, which are independent on the choice of the transformation.

Finally, we recall that, when defining quantum field theory on \ads,
some
caution should be taken with the \bc at infinity, since such \st is
not
globally hyperbolic [13]. In the following, as explained in detail
in [6],
we shall use "transparent" \bc, since they are more suitable for the
physical process at hand.

In ref. [6] we considered already this case, but we neglected the
difference in the global properties of \ADS and
\PDS: the answer we obtained was of course that no
Hawking radiation at all can be detected. This confirms the purely
topological nature of the Hawking radiation in this model.

The ground state for the Hawking radiation is given, if one excludes
from the theory the negative mass states, by the \ADS \st with metric
(3.11), while the positive mass \bh is described by a metric of
the form
(3.17).  In the following discussion, however, we shall define the
$\sigma^\pm$
coordinates in a different way from (3.18). This is possible
because the
$GL(2,R)$ group of \trans:
$$\xn\rightarrow {a\xn+b\over c\xn+d},\qquad\xm\rightarrow
{a\xm+b\over c\xm+d},
\eqno(4.1)$$
with $ad-bc\not=0$ is the isometry group of the metric (3.11), so
that one
can define different
coordinates which lead
to the same metric form: $GL(2,R)$ should therefore be considered
as the invariance group of the $ADS^0$ vacuum. Thus, we define
$\sigma^\pm$ such that:
$$\xm=\ab\e{\ba\sm},\qquad\qquad\xp=\ab\e{\ba\sp},\eqno(4.2a)$$
in $F$ and
$$\xm=-\ab\e{\ba\sm},\qquad\qquad\xp=-\ab\e{\ba\sp},\eqno (4.2b)$$
in $P$, where the $F$ and $P$ regions respectively correspond
to positive or negative $\xp$ and are displayed in fig. 5:
\PDS covers the $\xp\xm\ge 0$ region of \ADS and the \bh horizon
corresponds to
$\xp\xm=0$.

The physical meaning of the \ct (4.2) looks clearer if one uses
the standard
\st\coo:
$x=(\xm-\xp)/2,t=(\xm+\xp)/2,
\sigma=(\sm-\sp)/2,\tau=(\sm+\sp)/2.$
In these \coo, \ADS is given by:
$$\ds{1\over \lq x^2}(-dt^2+dx^2),\eqno(4.3)$$
with $-\infty<t<\infty,0<x<\infty$ and the \PDS \bh has
line element:
$$\ds{a^2\over\sinh^2(\ba\sigma)}(-d\tau^2+d\sigma^2),\eqno(4.4)$$
with $-\infty<\tau<\infty, 0< \sigma < \infty$.
The two metrics are related by the change of variable:
$$\eqalign{&x=\pm\ab\e{\ba\tau}\sinh(\ba\sigma),\cr
&t=\pm\ab\e{\ba\tau}\cosh(\ba\sigma).\cr}\eqno(4.5)$$

The change of variable (4.5), as already noticed, does not
correspond to
the motion of a physical observer, since its trajectory would be
space-like. By consequence, the causal structure is different from
that encountered in the Rindler problem. However, the mathematical
structure
is identical and for this reason we prefer to adopt these \coo instead
of the more intuitive ones defined by (3.18).

Let us consider the quantization of a single massless scalar field $f$
in the fixed background defined by $ADS^0$ and $ADS^+$.
In \ADS, $f$ can be expanded in terms of the
basis:
$$\ub_k=\nor\e{ikx-i\omega t},\eqno(4.6)$$
with $\omega=|k|$ and $-\infty<k<\infty$.
The $k>0$ modes are left-moving waves:
$$\nor\e{-i\omega\xp},$$
while $k<0$ corresponds to right-moving waves
$$\nor\e{-i\omega\xm}.$$
These two sets are positive frequency with respect to the Killing
vector
$\partial_t$.

Analogously, one can define a basis in the \bh regions $P$ and $F$:
\bigskip
$$\uf_k=\nor\e{ik\sigma-i\omega\tau}\qquad{\rm in}\ F;\qquad\qquad
0\qquad
{\rm in}\ P;\eqno(4.7a)$$
and
$$\up_k=\nor\e{ik\sigma+i\omega\tau}\qquad{\rm in}\ P;\qquad\qquad
0\qquad
{\rm in}\ F;\eqno(4.7b)$$
with $\omega=|k|$, which are positive frequency with respect to
$\partial_\tau$.

This basis can be  continued to the whole $x-t$ plane. However, due to
the change of sign at $\xp=\xm=0$, it is not analytic at that point.
Consequently, it defines
an alternative Fock space, which corresponds to a vacuum state
different
from that defined  by the basis (4.6). By a standard argument [14],
the
\PDS vacuum state will therefore appear to an observer in the
\ADS vacuum as
filled of thermal
radiation. The actual content can be obtained by calculating the
Bogoliubov
coefficients between the two vacua. To this end, it is useful to
define some
linear combinations of the \PDS basis:
$$\uf_k+\e{\pia}\up^*_{-k}\eqno(4.8a),$$
 and
$$\uf^*_{-k}+\e{\pia}\up_k.\eqno(4.8b)$$
Contrary to (4.7), this basis is analytic
for all real values of $\xp$ and $\xm$ and hence shares the same
vacuum
state with the \ADS basis (4.6). In fact (4.8a) is proportional to
$$(\ub_k)^{-i\iab}\qquad {\rm for}\ k>0,\qquad\qquad
(\vb_k)^{-i\iab}\qquad {\rm for}\ k<0,$$
while (4.8b) is proportional to
$$(\vb_k)^{i\iab}\qquad {\rm for}\ k>0,\qquad\qquad
(\ub_k)^{i\iab}\qquad\qquad {\rm for}\ k<0.$$

Comparing the expansion of $f$ in the two basis:
$$f=b_k\uno\up_k+ b_k\und\up_k^*+b_k\due\uf_k+b_k\dud\uf_k^*=$$
$$=\norm [d_k\uno \left(\e\iam\uf_k+\e{-\iam}\up^*_{-k}\right)+
\eqno(4.9)$$
$$+d_k\due \left(\e\iam\uf^*_{-k}+\e{-\iam}\up_k\right)+{\rm h.c.}],$$
where the operators $d_k^{(i)}$ annihilate the \ADS vacuum state
$|0_0>$,
while the $b_k^{(i)}$
annihilate the \bh vacuum state $|0_+>$, one gets
$$b_k\uno=\norm\left[\e\iam d_k\due+\e{-\iam}
d_k\und\right]$$
$$b_k\due=\norm\left[\e\iam d_k\uno+\e{-\iam}
d_k\dud\right]$$
from which one can read off the Bogoliubov coefficients.

In particular, an observer in the \ADS vacuum detects a thermal
flux of
particles with spectrum:
$$<_+0|d_k^\dagger d_k|0_+>={1\over\e\pia -1}.\eqno(4.10)$$
When integrated, it gives for the total flux of $f$-particle energy:
$$G={a^2\lq\over 48}.\eqno(4.11)$$
The flux corresponds to a Planck spectrum at temperature
$T_0={a\la\over 2\pi}$,
the Hawking temperature (3.6) of the \bh.
The local temperature at a given point is instead:
$$T=(g_{00})^{-\ha}T_0={\la\over 2\pi}\sinh(a\la\sigma)$$
which goes to zero at spatial infinity.
As a check of the previous result we can compute the outgoing
stress tensor
for the Hawking radiation
in terms of the relationship (3.18) between the coordinates
$\sigma$ and
$x$. In general  it is proportional to the Schwarzian derivative
of the
function $x^-=F(\sigma^-)$:
$$<0|T_{--}|0>= -{1\over 24}\biggl[{F''' \over F'}-{3\over 2}\biggl(
{ F''\over  F'}\biggr)^2\biggr],$$
where the primes  indicate derivation with respect to $\sigma^-$.
With $F$ given by (3.18) and making use of (3.3) we get
$$<0|T_{--}|0>= {1\over 48}a^2\la^2= {1\over 24}\pho M\la.
\eqno(4.12)$$
in accordance with eq. (4.11).
The stress energy tensor for the Hawking radiation has therefore
the constant (thermal) value which one would naively expect in view
of the specific heat expression (3.6) and coincides with the $k=0$
limit
of the result obtained in [6].
Of course, we are not considering the \br
of the metric, so that the flux is independent of time.
In the real physical
process the mass and hence the radiation rate, will decrease
as the \bh radiates until the total
evaporation.
This means that in this approximation,
loosing mass through the radiation the \bh will
settle down to its ground state $M=0$ which is represented by
$ADS^0$ vacuum.
It is interesting to note that the stress tensor of the Hawking
radiation
 has the same invariance group of the schwarzian derivative, i.e the
$GL(2,R)$ group realised as the fractional transformation (4.1),
which is also the invariance group of  $ADS^0$, as discussed before.

Let us now discuss the inclusion in our calculations of the back
reaction
of the radiation on the gravitational background.
We can guess in view of the general features of our model
that  this effect is not related to a change of the
geometry of the spacetime but to a change of the boundary conditions.
In fact the classical solutions of our theory are distinguishable
only by means of global properties and the Hawking radiation is
a purely topological effect.
The back reaction of the radiation on the metric can be studied in
a standard
way by considering, in the quantization of the scalar fields $f$,
the contribution of the trace anomaly to  the effective action.
This contribution is the well-known Polyakov-Liouville action
which in the conformal gauge is a local term.
The semiclassical action in the conformal gauge is
$$S={1\over\pi}\int d^2\sigma\left[\left( 2\dd\dm\rho +{\la^2\over 2}
e^{2\rho}\right)e^{-2\phi}
+{1\over2}\sum_{i=1}^N\dd f_i\dm f_i-{N\over 12}\dd \rho\dm\rho
\right].\eqno(4.13)$$
Where the first three terms come from the classical action (3.9)
whereas the fourth describes the trace anomaly.
The determinant of the kinetic energies in the $(\rho, \phi)$ space is
proportional to $\exp(-4\phi)$ so that there is no degeneration
of the
kinetic energies in the physical field space ($\exp(-2\phi)>0$).
This behaviour is to be compared with the CGHS dilaton gravity theory
where the kinetic energies become degenerate at the point of the
field space
where $\exp(-2\phi)=N/12$ [15].
The  ensuing equation of motion and constraint  are
$$\eqalign{\dd\dm f_i&=0,\cr \dd\dm\rho&=-{\la^2\over4}e^{2\rho},\cr
\dd\dm\ed&= {\la^2\over2}e^{2\rho}\biggl({N\over 24}-\ed\biggr),\cr
 \dd^2\ed &-2\dd\rho\dd\ed=-{1\over 2}\sum_{i=1}^N\dd f_i\dd f_i+
{N\over 12}\biggl[(\dd\rho)^2-\dd^2\rho +t_+\biggr],\cr
 \dm^2\ed &-2\dm\rho\dm\ed=-{1\over 2}\sum_{i=1}^N\dd f_i\dd f_i
+{N\over 12}\biggl[(\dm\rho)^2-\dm^2\rho +t_-\biggr],\cr}
\eqno(4.14)$$
where the functions $t_+,t_-$ must be determined using boundary
conditions.
The  semiclassical equations of motion differ from
the classical ones just in the shift
$$\exp(-2\phi)\to \exp(-2\phi)+N/24\eqno(4.15)$$
of the dilaton and for the presence in the constraint equation
of the term stemming from the conformal anomaly.
This result can be achieved
directly from the action (4.13)
by noting that the term describing the conformal anomaly
can be reabsorbed in the classical action by means of the former
redefinition of the dilaton.
This means that the solutions of the semiclassical equations of
motion are locally the same as the classical ones and differ
from those only through the boundary conditions.

If one ignores the presence of the
boundary on the spacetime along the line where $\exp(-2\phi)$
becomes negative,
i.e.  if one  takes maximally extended solutions there is, as seen
previously,
no Hawking radiation and therefore no back reaction.
On the other hand when the boundary is present and supported by
appropriate  boundary conditions
the problem of the back reaction can in principle be analysed
by studying, together with eq.(4.14), the dynamical evolution
of the boundary
in the context of the semiclassical theory.
We will not address the problem in this general setting, but will
study it in a
static approximation which is however consistent with our previous
treatment of the Hawking radiation process.
Let us consider the process in which the \bh radiates as a
succession of static
states with decreasing mass. Being the solutions static they
will depend only on the variable $\sigma=\sigma^--\sigma^+$.

The equations of motion and the constraint (4.14),
with $f_i=0$, become now:
$$\eqalign{\rho''&={\lq\over 4}e^{2\rho},\cr
\psi''&={\lq\over 2}e^{2\rho}\biggl(\psi-{N\over 24}\biggr),\cr
\psi''&-2\rho'\psi'={N\over 12}[(\rho')^2-\rho''],\cr}\eqno(4.16)$$
where $\psi=\exp(-2\phi)$ and the primes represent derivatives
with respect to
$\sigma$.
Integration of the equations of motion (4.16) gives two class of
solution
(we do not consider the solutions which correspond to $ADS^-$)
$$\eqalign{e^{2\rho}&={4c^2\over \lq}\sinh^{-2}(c\sigma),\cr
\ed&=\biggl(e^{-2\phi_0}+{N\over 24}c \sigma\biggr)\tanh^{-1}
(c\sigma),\cr}
\eqno(4.17)$$
and
 $$\eqalign{e^{2\rho}&={4\over \lq\sigma^2},\cr
\ed&={e^{-2\phi_0}\over \sigma}+{N\over 24}.\cr}
\eqno(4.18)$$
In eq. (4.17) $c$ is an integration constant, for $ c=a\la/2$
the metric part of the
solution coincide with the classical solution (3.17),
it describes therefore $ADS^+$.
The solution (4.18) can be considered the ground state of the
semiclassical
theory and coincides with the ground state of the classical theory
($ADS^0$) after the shift (4.15) of the  dilaton.
Let us now discuss the semiclassical solutions (4.17), (4.18).
First we note that for $e^{2\phi}<<24/N$ the semiclassical
solutions (4.17),(4.18)
behave as the classical ones (3.11),(3.17). Hence in the weak coupling
regime
we can safely ignore the  back reaction.
Moreover, one can easily realize that the back reaction
affects only the dilaton but not the metric. This is again a feature
which is connected with the purely topological nature of the  Hawking
radiation.
The states with decreasing values of $c$ or equivalently of the
\bh mass,
in eq (4.17) describe the evolution of the \bh  when the
evaporation process takes
place if we think of it as a succession of static states.
In particular, the limit $c \to 0$ ($M\to 0$) will tell us what
is the end of the evaporation process.
Performing the limit $c\to 0$ in the solution (4.17) we get the
vacuum solution (4.18). Thus the end point of the evaporation
process is exactly the $ADS^0$.
In a pictorial description of the process we see that the region
of the $ADS^0$ spacetime covered by the coordinate $\sigma$ increases
(equivalently the horizon of the \bh recedes) as the
\bh looses mass through the radiation and, at the end point,
this region coincides with the whole $ADS^0$ space (see figure 6).

\beginsection 5. Summary and outlook.
\smallskip
We have studied the \4d analog of the scalar-gravity JT theory and
shown that it admits regular, \af magnetically charged solutions
which can be interpreted as non-singular \bhs.

The final state of the evaporation
process of these \bhs is given, as usual, by the \el of the metric.
The evaporation can be described by means of a \2d effective theory
which is governed by the JT action. This theory admits \ads solutions
which can be interpreted as the conformal vacuum or as \bhs depending
on the \bc one chooses. These are determined, trough the form of
the dilaton, by the \4d
physics we want to describe with the \2d model.

We have calculated the flux of Hawking radiation in a way analogous to
that used in the case of Rindler \st and also by computing the
energy-momentum tensor. Moreover, by means of a static approximation,
we have discussed the inclusion of the \br of the fields to the
radiation, arguing  that it results in a smooth evolution of the \bh
towards the \gs, in contrast with the CGHS case , where some
singularities appear in the evolution.

A more careful analysis should involve the dynamical evolution of the
\bc. It is not clear at the moment whether
this can be deduced from the
lagrangian or, owing to the global nature of the problem,
some further input should be added to the formalism.
\smallskip

{\bf Acknowledgment}

{\noindent This work was partially supported by MURST.}
\smallskip
\beginref

\ref [1] C.G. Callan, S.B. Giddings, J.A. Harvey and A. Strominger,
\PRD {\bf D45}, 1005 (1992).

\ref [2] E. Witten, \PRD {\bf 44}, 314 (1991);
G. Mandal, A.M. Sengupta and S.R. Wadia, \MPLA  {\bf 6}, 1685 (1991);
R.B. Mann, A. Shiekh and I. Tarasov, \NPB {\bf 341}, 134 (1990);
R.B. Mann, Gen. Rel. Grav. {\bf 24}, 433 (1992);
Y. Frolov, \PRD {\bf 46}, 5383 (1992);
J.P.L. Lemos and P.M. S\'a, \PRD {\bf 49}, 2897 (1994);
T. Banks, A. Dabholkar,  M. R. Douglas and M. O'Loughlin, \PRD {\bf
45}, 3607 (1992);
L. Susskind and L. Thorlacius, \NPB {\bf 382}, 123 (1992);
S.B. Giddings and W.M. Nelson, \PRD  {\bf 46}, 2486 (1992).

\ref [3] D. Garfinkle, G.T. Horowitz and A. Strominger, \PRD {\bf 43},
3140 (1991);

\ref [4] G.W. Gibbons and K. Maeda, \NPB {\bf 298}, 741 (1988);
S.B. Giddings and A. Strominger, \PRD {\bf 46}, 627 (1992).

\ref [5] M. Cadoni and S. Mignemi, \PRD {\bf 48}, 5536 (1993).

\ref [6] M. Cadoni and S. Mignemi, \NPB (in press), hep-th 9312171.

\ref [7] C. Teitelboim, in {\sl Quantum Theory of gravity },
S.M. Christensen,
ed. (Adam Hilger, Bristol, 1984); R. Jackiw, {\sl ibidem}.

\ref [8] A. Ach\'ucarro and M.E. Ortiz, \PRD {\bf 48}, 3600 (1993).

\ref [9] M. Ba\~nados, C. Teitelboim and J. Zanelli,
\PRL {\bf 69}, 1849 (1992).

\ref [10] D. Christensen and R.B. Mann, \CQG {\bf 9}, 1769 (1992);
J.P.L. Lemos and P.M. S\'a, Preprint DF/IST-8.93.

\ref [11] J. Jackiw, Lectures given at the {\sl International
Colloquium on  Group Theoretical Methods in Physics},
Salamanca, June 1992.

\ref [12] E. Verlinde and H. Verlinde \NPB {\bf 406}, 43 (1993);
T.D. Chung, H. Verlinde \NPB {\bf 418}, 305 (1994).

\ref [13] S.J. Avis, C.J. Isham and D. Storey,
\PRD {\bf 18}, 3565 (1978).

\ref [14] N.D. Birrell and P.C.W. Davies, {\sl Quantum fields in
curved space}
(Cambridge Un. Press, 1982).

\ref [15] J.G. Russo, L. Susskind and L. Thorlacius \PLB {\bf 292},
13 (1992).

\endref
\end